\documentclass[conference,a4paper]{IEEEtran}
\ifCLASSINFOpdf
  \usepackage[pdftex]{graphicx}
  \graphicspath{{./}}
  \DeclareGraphicsExtensions{.pdf,.jpeg,.png}
\else
  \usepackage[dvips]{graphicx}
  \graphicspath{{./}}
  \DeclareGraphicsExtensions{.eps}
\fi
\usepackage[cmex10]{amsmath}
\newtheorem{proof}{Proof}
\newtheorem{lemma}{Lemma}

\usepackage{array}
\usepackage{mathrsfs}
\usepackage{amsfonts}
\usepackage{amssymb}

\newenvironment{myindentpar}[1]
{\begin{list}{}%
         {\setlength{\leftmargin}{#1}}
           \item[]%
}
{\end{list}}

\begin{document}
%\title{Trellis-Based Channel Estimation For Asynchronous IDMA With Superimposed Pilots}
\title{Turbo-AI: Iterative Machine Learning Based Channel Estimation for 2D Massive Arrays}

\author{\IEEEauthorblockN{Yejian Chen, Jafar Mohammadi, Stefan Wesemann and Thorsten Wild}
\IEEEauthorblockA{Bell Laboratories, Nokia, Lorenzstra{\ss}e 10, D-70435 Stuttgart, Germany\\
Yejian.Chen@Nokia-Bell-Labs.com}}


%Telephone: (800) 555--1212, Fax: (888) 555--1212}


\maketitle

\begin{abstract}
%\boldmath
Recently, Machine Learning (ML) is recognized as an effective tool for wireless communications and plays an evolutionary role to enhance Physical Layer (PHY) of 5th Generation (5G) and Beyond 5G (B5G) systems. In this paper, we focus on the ML-based channel estimation for 2-Dimensional (2D) massive antenna arrays. Due to the extremely high computational requirement for 2D arrays with \emph{Ordinary Training}, we exploit 2D Kronecker covariance model to perform \emph{Subspace Training} for vertical and horizontal spatial domain independently, which achieves a complexity cost saving factor $\mathcal{O}(M^4N^4)/\mathcal{O}(MN^4+NM^4)$ for ML with an $M\times N$ 2D-array. Furthermore, we propose an iterative training approach, referred to as \emph{Turbo-AI}. Along with \emph{Subspace Training}, the new approach can monotonically reduce the effective variance of additive noise of the observation, and update the Neural Network (NN) models by re-training. Furthermore, we propose a concept, named \emph{Universal Training}. It allows to use one NN for a wide range of Signal-to-Noise-Ratio (SNR) operation points and spatial angles, which can greatly simplify \emph{Turbo-AI} usage. Numerical results exhibit that \emph{Turbo-AI} can tightly approach the \emph{genie-aided} channel estimation bound, especially at low SNR.
%\\
\end{abstract}

%\begin{IEEEkeywords}
%Multiple-Input Multiple-Output (MIMO), Zero-Forcing (ZF), Maximum \emph{a posteriori} Probability (APP) detection, Overlapped 
%Subspace Detection (OSD).
%\end{IEEEkeywords}

% \fi
\IEEEpeerreviewmaketitle

%\vspace{1mm}
\section{Introduction}
\vspace{1mm}
Witnessing so many Artificial Intelligence (AI) inspired inventions and applications, no one is going to doubt, whether the AI techniques are changing our world. The recent rapidly developed AI techniques exhibit high potential for many technical challenges. In 2016 and 2017, AlphaGo, the AI program developed by DeepMind, defeated the human professional world champions twice in the most challenging strategy board game \emph{Go}, which can be recognized as a revolutionary landmark of deep learning \cite{ref2.00}--\cite{ref3.00}. In communications society, AI is attracting enormous attentions from both industry and academia for enhancing the new concepts and novel technologies within the context of 5th Generation (5G) New Radio (NR) \cite{ref4.00} and even opening new perspectives for Beyond 5G (B5G) systems. In this paper, we focus on the exploitation of AI for PHYsical (PHY) layer of wireless communications. It is well known that the performance of the key components of PHY layer are mathematically bounded by derivable optimums. Among them, the fundamental and the famous one is the Shannon limit \cite{ref5.00}. Obviously, it is impossible to outperform the optimum bound by means of deep learning. Instead, AI introduces an alternative path to approach the optimum bound with low complexity. Thus, AI provides us additional degrees of freedom to trade off the performance and complexity in PHY layer design for 5G and B5G system. There are potential AI applications for PHY layer, enumerated in \cite{ref6.00}, such as blind channel decoding and data detection, modulation recognition, channel estimation, and many others. 

Hence, our paper is motivated in this context by introducing AI for channel estimation to enhance PHY layer performance. Precise channel estimation belongs to one of the key technical prerequisites to support PHY layer operations, such as equalization and data detection, power control, precise Channel State Information (CSI) for transmit precoding and feedback, user pairing for Multiple-Input Multiple-Output (MIMO)  and Non-Orthogonal Multiple Access (NOMA) system, and so on. In \cite{ref7.00}, a Neural Network (NN) based channel estimation approach is proposed, in which the conventional Minimum Mean Square Error (MMSE) channel estimator is equivalently represented  under certain array-specific property assumptions as a two-layer NN. The conventional channel estimators, e.g. \cite{ref8.00}, are usually more complex than the Machine Learning (ML) based solutions \cite{ref9.00} for similar performance. Especially, the computational requirement will be significantly increased for massive Multiple-Input Multiple-Output (mMIMO) system \cite{ref10.00} in these conventional channel estimators with a large number of antenna elements. A proof-of-concept is provided in \cite{ref11.00} provides for NN-based channel estimation, by means of real-time experiment and measurement. In \cite{ref12.00}, the ML-based channel estimation is considered for mMIMO with different antenna configurations. 

In this paper, we will focus on the low-cost ML-based channel estimation with massive 2-Dimensional (2D) antenna panels, by exploiting the approximation in \cite{ref13.00}. With the Kronecker channel model, the spatial covariance matrix of a 2D-array can be decomposed as the Kronecker product of a vertical covariance matrix and a horizontal covariance matrix, respectively. The training for the full spatial covariance matrix of the 2D-array, denoted as highly complex \emph{Ordinary Training}, can be replaced by low complex \emph{Subspace Training} with two independent NNs in horizontal and vertical domains, respectively. Different combining strategies for channel estimates, obtained after the \emph{Subspace Training} in vertical and horizontal subspaces, can improve the channel estimation with reduced complexity. Furthermore, after combining the estimates from independent subspaces, the estimation error, which can be interpreted as additive noise, is still a Gaussian random variable, but with reduced variance. Let the initial observation be replaced by the less noisy channel estimates, and repeat the \emph{Subspace Training} and combining, to approach the optimum performance iteratively. Throughout this paper, this procedure is referred to as \emph{Turbo-AI}. It can be demonstrated through link level simulation that the \emph{genie-aided} bound can be tightly reached. Furthermore, we propose the concept \emph{Universal Training} to enhance the robustness of \emph{Turbo-AI} for practical implementation. 

This paper is organized as follows. In Section II we focus on the system model. An overview of \emph{Subspace Training}, \emph{Turbo-AI} and \emph{Universal Training} will be presented in Section III. In Section IV, numerical results of link layer performance are provided. Finally, Section V renders some conclusions.

\section{System Model}
As depicted in Fig.~1, let us consider a 2D-array system with $M$ vertical elements and $N$ horizontal elements. The samples $k = 1, \cdots, K$ are collected from time domain as temporal symbols or from frequency domain as subcarriers. Hence, the Single-Input Multiple-Output (SIMO) system can be represented by following equation as
% Equation 1
\vspace{1mm}
\begin{equation}
\label{eq1}
\mathbf{Y}_k = \mathbf{H}_k + \mathbf{Z}_k \textrm{,}
\vspace{1mm}
\end{equation}where the $M \times N$ matrix $\mathbf{Y}_k$ and $\mathbf{H}_k$ stands for the observations and wireless channel, respectively. The matrix $\mathbf{Z}_k$ denotes the Additive White Gaussian Noise (AWGN), element-wise characterized by noise power spectral density $N_0 = 2 \sigma^2$.
% Figure 1
\begin{figure}[!htp]
\vspace{-16mm}
\centering
\includegraphics[width=3.5in]{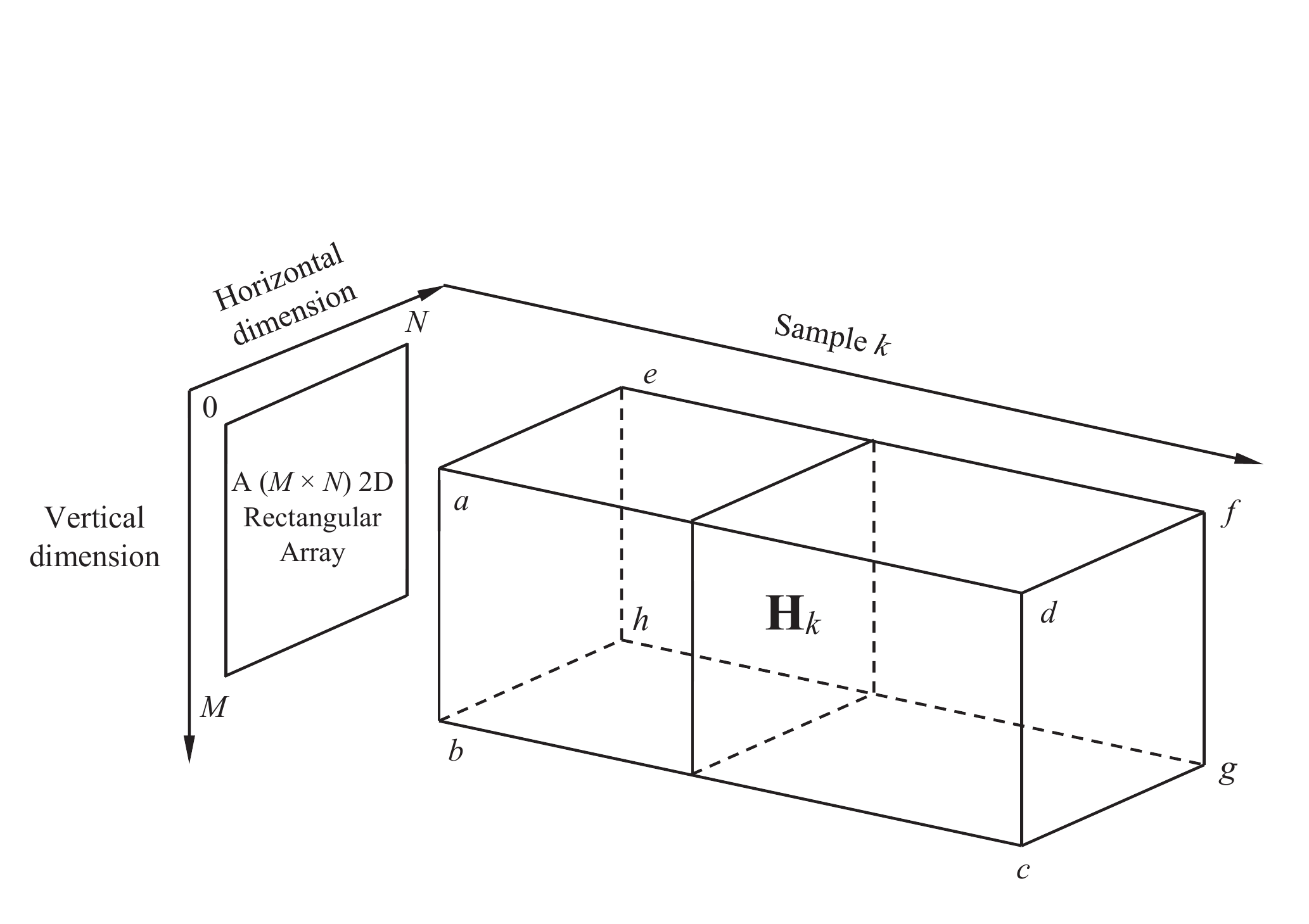}
\vspace{-10mm}
\caption{Signal model with a 2D antenna panel}
\vspace{-1mm}
\end{figure}
In \cite{ref13.00}, it is exhibited that the covariance matrix $\mathbf{R}$ of a 2D-array can be approximated as
% Equation 2
\vspace{1mm}
\begin{equation}
\label{eq2}
\mathbf{R} \approx \mathbf{R}_h \otimes \mathbf{R}_v  \textrm{,}
\vspace{1mm}
\end{equation}where $\mathbf{R}_h$ and $\mathbf{R}_v$ denote the common horizontal and vertical covariance matrix in the sense of 1D-arrays in row and column, respectively. The covariance matrices $\mathbf{R}_v$, $\mathbf{R}_h$ and $\mathbf{R}$ are thus $M \times M$, $N \times N$ and $MN \times MN$ matrices. The spatial correlation between $i$-th and $j$-th antenna in $\mathbf{R}_h$ \cite{ref14.00}, denoted as $\mathbf{R}_{h}^{(i,j)}$, can be computed as
% Equation 3
\vspace{1mm}
\begin{equation}
\label{eq3}
\mathbf{R}_{h}^{(i,j)} = \frac{1}{\sqrt{2}\varrho}\int\limits_{-\pi}^{\pi}\exp\Big[ -\frac{\sqrt{2} |\phi|}{\varrho} + j\frac{2\pi}{\lambda} \Delta d_{i,j} \sin(\theta+\phi) \Big] \textrm{d}\phi \textrm{,}
\vspace{2mm}
\end{equation}where $\Delta d_{i,j}$ denotes the distance between both antenna elements $i$ and $j$. Parameters $\varrho$ and $\theta$ stand for the angular spread and Direction of Arrival (DoA), respectively. The given random variable $ \phi$ obeys Laplacian power angular distribution concerning $\varrho$ and $\theta$. Equation (3) is also valid for vertical spatial domain $\mathbf{R}_v^{(i,j)}$. Considering the independent and identically distributed (i.i.d.) Rayleigh fading channel, it holds
% Equation 4
\vspace{1mm}
\begin{equation}
\label{eq4}
\textrm{vec}(\mathbf{H}_{k}) = \mathbf{R}^{0.5} \textrm{vec}(\mathbf{H}_{w,k})  \textrm{.}
\vspace{1mm}
\end{equation}
The elements in the $M \times N$ matrix $\mathbf{H}_{w,k}$ follow i.i.d. Rayleigh distribution with $E[|\mathbf{H}_{w,k}|^2]=1$, representing the sample $k$ randomly selected from time or frequency domain. Thus, the channel matrix $\mathbf{H}_k$ can be equivalently computed as
% Equation 5
\vspace{1mm}
\begin{equation}
\label{eq5}
\mathbf{H}_{k} = \mathbf{R}^{0.5}_v \mathbf{H}_{w,k} (\mathbf{R}^{0.5}_h)^T \textrm{.}
\vspace{1mm}
\end{equation}

\section{Channel Estimation and Machine Learning}
In this section, let us focus on the channel estimation and training for machine learning. First of all, let us discuss about the benchmark solution of channel estimation. For a linear signal model in presence of Gaussian noise, the  Linear Minimum Mean Square Error (LMMSE) channel estimator is regarded as the optimal linear estimator, in sense of being capable of minimizing the Mean Square Error (MSE). For a 2D-array system, depicted in Fig.~1, the MMSE weight matrix can be computed as 
% Equation 6
\vspace{1mm}
\begin{equation}
\label{eq6}
\mathbf{W}_\textrm{genie} = \mathbf{R} (\mathbf{R} + \sigma^2 \mathbf{I}_{MN})^{-1} \textrm{.}
\vspace{1mm}
\end{equation}If the covariance matrix $\mathbf{R}$ is \emph{a priori} known, it is referred to as \emph{genie-aided} solution. Nevertheless, the drawback is obvious as well: if the number of antenna elements $M$ and $N$ become large, the matrix inversion in (6) can be computationally intensive. 
\subsection{Estimation Through Subspaces}
Let us re-consider the problem from the view point of horizontal and vertical subspaces as depicted in Fig.~2. We can stack the data as slices horizontally and vertically to create two observations. Let us define the $N \times M$ matrix $\mathbf{H}_h$ and $M \times N$ matrix $\mathbf{H}_v$ as
% Equation 7
\vspace{1mm}
\begin{equation}
\label{eq7}
\mathbf{H}_h = \big[ \mathbf{h}_1^{(h)} \cdots \mathbf{h}_M^{(h)} \big]
\vspace{1mm}
\end{equation}
% Equation 8
\vspace{-5mm}
\begin{equation}
\label{eq8}
\mathbf{H}_v = \big[ \mathbf{h}_1^{(v)} \cdots \mathbf{h}_N^{(v)} \big] \textrm{,}
\vspace{2mm}
\end{equation}where $M$ samples of $N\times1$ vector $\mathbf{h}_m^{(h)}$ and $N$ samples of $M\times1$ vector $\mathbf{h}_n^{(v)}$ are obtained in horizontal and vertical subspaces, respectively. Consider MMSE channel estimation for both subspaces individually, it holds
% Equation 9
\vspace{1mm}
\begin{equation}
\label{eq9}
 \hat{\mathbf{H}}_h =  \mathbf{W}_h \mathbf{Y}^T=\mathbf{R}_h (\mathbf{R}_h + \sigma^2 \mathbf{I}_{N})^{-1}\mathbf{Y}^T
\vspace{1mm}
\end{equation}
% Equation 10
\vspace{-5mm}
\begin{equation}
\label{eq10}
\hat{\mathbf{H}}_v = \mathbf{W}_v \mathbf{Y}=\mathbf{R}_v (\mathbf{R}_v + \sigma^2 \mathbf{I}_{M})^{-1}\mathbf{Y}
\vspace{2mm}
\end{equation}where $\mathbf{W}_h$ and $\mathbf{W}_v$ denote $N \times N$ and $M \times M$ MMSE weighting matrices, and the observation is denoted as $M \times N$ matrix $\mathbf{Y}$. Let us simply consider the arithmetic mean to combine the estimates from vertical and horizontal subspaces. It holds 
% Equation 11
\vspace{2mm}
\begin{equation}
\label{eq11}
 \mathbf{\hat{H}}_a = 0.5(\hat{\mathbf{H}}_h^T + \hat{\mathbf{H}}_v) =0.5(\mathbf{Y}\mathbf{W}_{h}^{T}+\mathbf{W}_v \mathbf{Y}) \textrm{.}
\vspace{2mm}
\end{equation}The result in (11) can be further reformulated as a vectorized expression as
% Equation 12
\vspace{2mm}
\begin{equation}
\label{eq12}
\textrm{vec}(\mathbf{\hat{H}}_a) = 0.5(\mathbf{I}_N \otimes \mathbf{W}_v +  \mathbf{W}_h \otimes \mathbf{I}_M) \textrm{vec}(\mathbf{Y})\textrm{,}
\vspace{2mm}
\end{equation}where the $MN \times MN$ matrix $\mathbf{W}_a = 0.5(\mathbf{I}_N \otimes \mathbf{W}_v +  \mathbf{W}_h \otimes \mathbf{I}_M)$ is the effective weighting with respect to the subspace combining with arithmetic mean. Similarly, we can exploit the geometric mean to perform the subspace combining. It holds
% Equation 13
\vspace{2mm}
\begin{align}
\textrm{vec}(\mathbf{\hat{H}}_g) & = (\mathbf{I}_N \otimes \mathbf{W}_v)^{0.5}(\mathbf{W}_h \otimes \mathbf{I}_M)^{0.5} \textrm{vec}(\mathbf{Y}) \\
 & = (\mathbf{W}_h^{0.5} \otimes \mathbf{W}_v^{0.5})\textrm{vec}(\mathbf{Y}) \textrm{,} \nonumber
\end{align}where the $MN \times MN$ matrix $\mathbf{W}_g = \mathbf{W}_h^{0.5} \otimes \mathbf{W}_v^{0.5}$ is the effective weighting with geometric mean. With the properties of Kronecker product, it yields,
% Equation 14
\vspace{2mm}
\begin{equation}
\label{eq14}
\mathbf{\hat{H}}_g = \mathbf{W}_v^{0.5} \mathbf{Y} (\mathbf{W}_h^{0.5})^T\textrm{.}
\vspace{2mm}
\end{equation}
%Notice that the total additive noise in the vertical and horizontal domains is obviously the same data. Being split and reshuffled into the two subspaces, the input vectors of horizontal NN is independent from any individual input vectors of vertical NN. The AWGN realizations of two observations turn out to be quasi-independent, referring to the marks with different colors in Fig. 2. 
Notice that the same data is reordered as the inputs for the horizontal NN and vertical NN, respectively. Nevertheless, the reshuffling patterns of corresponding AWGN components within both observations exhibit strong independence, represented as colorful spots in Fig.~2 for examples.
% Figure 2
\begin{figure}[!htp]
\vspace{-16mm}
\centering
\includegraphics[width=3.6in]{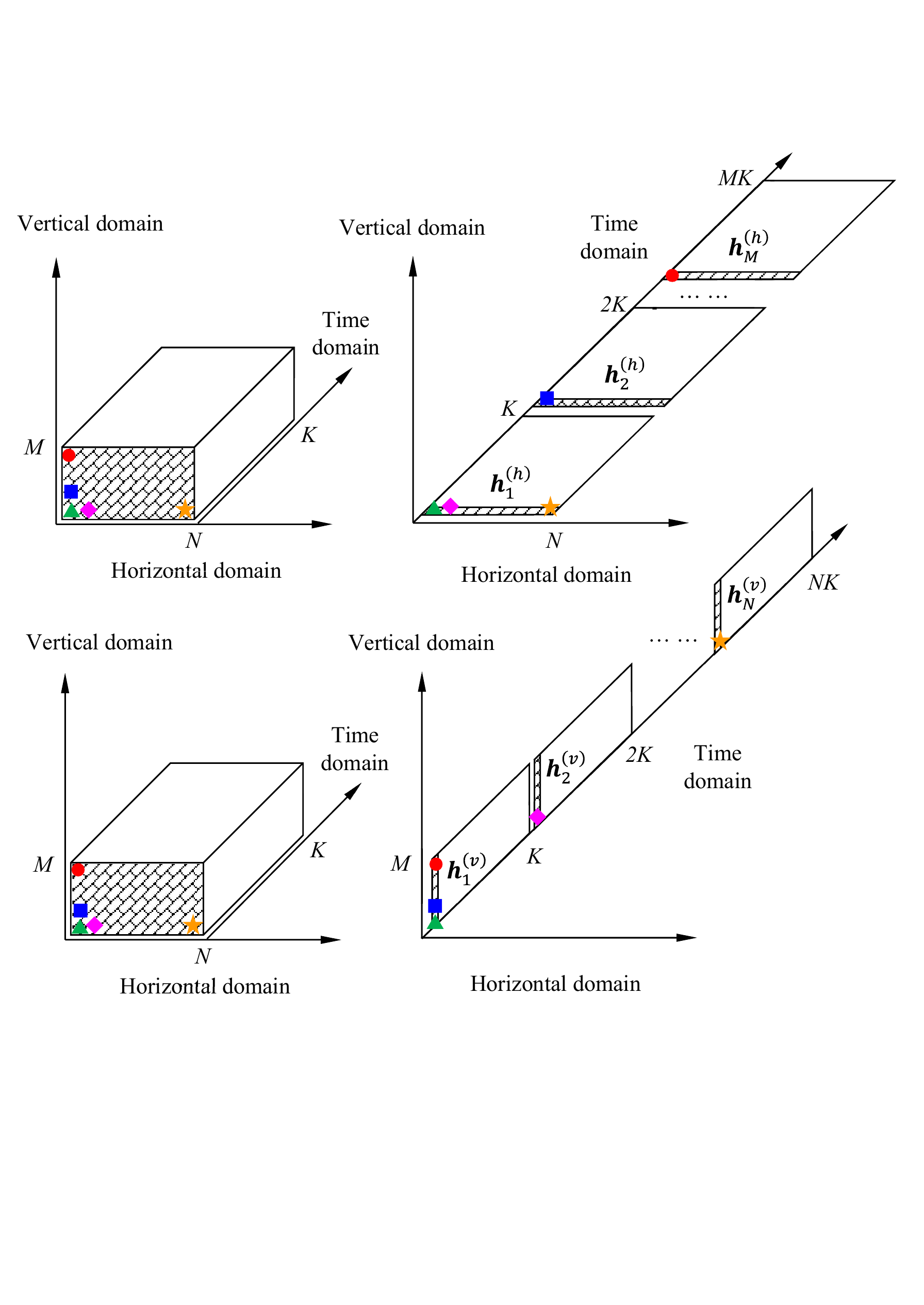}
\vspace{-37mm}
\caption{Stacking $M$ vertical slices in horizontal spatial domain and stacking $N$ horizontal slices in vertical spatial domain}
\vspace{-2mm}
\end{figure}The gain from combining the outputs of two NNs after \emph{Subspace Training} comes from the fact that the output randomness of each estimated element is originated from training these independent elements. It can be exhibited that  the variance post-processing noise can be reduced after subspace combining, e.g. arithmetic mean. Let $\textrm{VAR}[\mathbf{Z}_a^{(i,j)}]$ the variance of the AWGN component on $i$-th row and $j$-th column in matrix $\mathbf{Z}_a$ after exploiting arithmetic mean. It holds, 
% Equation 15
\vspace{1mm}
\begin{equation}
\label{eq15}
\frac{\rho}{4}N_0 < \textrm{VAR}\big[ \mathbf{Z}_a^{(i,j)}\big] \leq N_0 
\vspace{1mm}
\end{equation}
with $\rho=\sum_{m=1}^{M} E[|\mathbf{W}_v^{(i,m)}|^2]+\sum_{n=1}^{N} E[|\mathbf{W}_h^{(j,n)}|^2]$ and $0 \leq\rho\leq 2$. The variance of the original AWGN component is $\textrm{VAR}[\mathbf{Z}^{(i,j)}]=N_0=2\sigma^2$. The proof of (15) will be provided in \emph{Appendix A}. 
 
In order to explore the property of arithmetic mean and geometric mean, let us introduce the notations $\mathbf{y} = \textrm{vec}(\mathbf{Y})$, $\mathbf{A} = \mathbf{I}_N \otimes \mathbf{W}_v$ and $\mathbf{B} = \mathbf{W}_h \otimes \mathbf{I}_M$. With the arithmetic mean, the deviation of the channel estimates from the \emph{genie-aided} MMSE channel estimator can be computed as
% Equation 16
\vspace{1mm}
\begin{equation}
\label{eq16}
D_a = \frac{1}{MN}\mathbf{y}^H\big[ \frac{1}{2}(\mathbf{A}+\mathbf{B})-\mathbf{W}_\textrm{genie}\big]^2\mathbf{y}
%D_a = \frac{1}{MN}\mathbf{y}^H\big[ \frac{1}{2}(\mathbf{I}_N \otimes \mathbf{W}_v+\mathbf{W}_h \otimes \mathbf{I}_M)-\mathbf{W}_\textrm{genie}\big]^2\mathbf{y}
\vspace{1mm}
\end{equation}and with geometric mean, the deviation can be similarly computed as
% Equation 17
\vspace{1mm}
\begin{equation}
\label{eq17}
D_g = \frac{1}{MN}\mathbf{y}^H(\mathbf{A}^{0.5}\mathbf{B}^{0.5}-\mathbf{W}_\textrm{genie})^2\mathbf{y} \textrm{.}
%D_g = \frac{1}{MN}\mathbf{y}^H\big[(\mathbf{I}_N \otimes \mathbf{W}_v)^{0.5}(\mathbf{W}_h \otimes \mathbf{I}_M)^{0.5}-\mathbf{W}_\textrm{genie}\big]^2\mathbf{y} \textrm{.}
\vspace{1mm}
\end{equation}Furthermore, it holds
% Equation 18
\vspace{1mm}
\begin{equation}
\label{eq18}
D_a - D_g =  \frac{1}{MN}\mathbf{y}^H\mathbf{L}(\mathbf{P}+\mathbf{Q}) \mathbf{y}
\vspace{1mm}
\end{equation}with $\mathbf{L}=(\mathbf{A}^{0.5}-\mathbf{B}^{0.5})^2$, $\mathbf{P}=\mathbf{A}+\mathbf{B}-2\mathbf{W}_\textrm{genie}$ and $\mathbf{Q}=2(\mathbf{A}^{0.5}\mathbf{B}^{0.5}-\mathbf{W}_\textrm{genie})$. According to our computational analysis for our dataset with both i.i.d. Rayleigh channel assumption and 3GPP compliant channel model, the following holds
% Equation 19
\vspace{0mm}
\begin{equation}
\label{eq19}
D_a \geq D_g \textrm{,}
\vspace{1mm}
\end{equation}which indicates that the geometric mean can provide the channel estimates with better quality. For computing the geometric mean $\hat{\mathbf{H}}_g$, the matrix square root operation $(\cdot)^{0.5}$ has to be invoked for matrices $\mathbf{W}_v$ and $\mathbf{W}_h$, which introduces additional complexity due to the non-linear operation.
 
\subsection{Neural Network and Complexity}
It is exhibited in \cite{ref7.00} that a MMSE channel estimator can be approximated as a two-layer NN network. Thus, in a $M \times N$ 2D-array system, with a $MN \times MN$ spatial covariance matrix $\mathbf{R}$, the number of input parameters is $2 M^2N^2$ real values, if the sample covariance matrices are exploited for the training. Obviously, the complexity will be significantly increased, if a massive 2D-array is deployed. In \cite{ref7.00}, matrix transformation and decomposition are considered to reduce the number of NN parameters for antenna array with special structure. With the analysis in the previous subsection, the Kronecker model allows us to exploit the spatial decomposition naturally,  and process the horizontal and vertical subspaces individually as Uniform Linear Array (ULA) antennas. As depicted in Fig.~4, we exploit two NNs in parallel, each of which has simple structure as a two-layer network with ReLu activation. The same $M \times N \times K$ data structure, see Fig.~2, will be reconstructed and delivered to both NNs as the observations for vertical and horizontal subspaces. The input and the output of the NNs are spatial sample covariance matrices $\mathbf{R}_v$, $\mathbf{R}_h$ and the weight matrices $\mathbf{W}_v$, $\mathbf{W}_h$, respectively. Fig.~3 exhibits how to perform the combining for vertical and horizontal subspaces in sense of arithmetic mean and geometric mean, according to the analysis in the previous subsection. 
% Figure 3
\begin{figure}[!htp]
\label{fig:fig3}
\centering
\vspace{-18mm}
\includegraphics[width=3.6in]{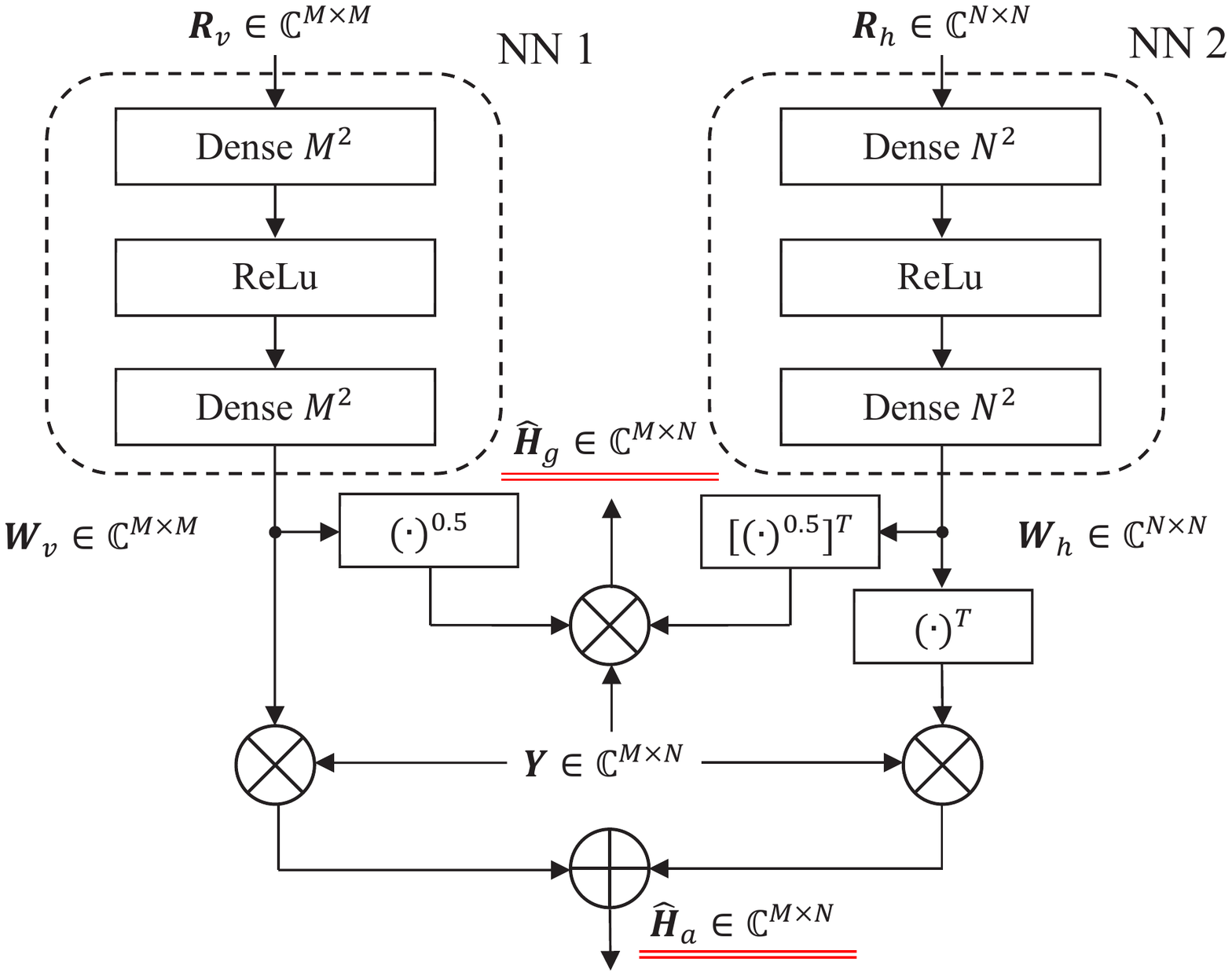}
\vspace{-51mm}
\caption{Two parallel NNs for \emph{Subspace Training}}
\vspace{-2mm}
\end{figure}Exploiting the full MMSE \emph{Ordinary Training} for the $M \times N$ 2D-array, two concatenated $(MN)^2$-In-$(MN)^2$-Out dense layers in one NN have to be considered. The number of real values as the input parameters is $2M^2N^2$. The number of the stored real values of both dense layers is $8M^4N^4$. Considering the cost of parameter passing, ReLu operation and real/imaginary part matrix multiplication, the complexity of the full MMSE \emph{Ordinary Training} is $\mathcal{O}(M^4N^4)$. In Fig.~3, two parallel NNs are depicted, each of which has two concatenated dense layers. The number of real values as the input parameters of the network is $2(M^2 + N^2)$. The number of stored real values of both NNs is $4(M^4+N^4)$. The total cost of \emph{Subspace Training} to involve parameter passing, ReLu operation, real/imaginary part matrix multiplication and subspace combining is $\mathcal{O}(MN^4+NM^4)$. Let us define a cost saving function $\eta(M,N)$ as
% Equation 20
\vspace{1mm}
\begin{equation}
\label{eq20}
\eta(M,N) \approx \frac{\mathcal{O}(M^4N^4 )}{\mathcal{O}(MN^4+NM^4)} \textrm{.}
\vspace{1mm}
\end{equation}For a $(M=4,N=8)$ and a $(M=8,N=16)$ 2D-array systems, the cost saving coefficients with \emph{Subspace Training} are approximately 50 and 450, respectively. 

\subsection{Turbo-AI}
\emph{Turbo-AI} is an iterative procedure, which consists of an iterative training stage and an iterative estimation stage. Let us explore the operation principle of \emph{Turbo-AI} in this subsection. After being combined in horizontal and vertical subspaces, the channel estimates, obtained by the NN described in previous subsection, still exhibit residual additive noise, which is Gaussian, but with a reduced variance, compared to the noise in the channel observation input. 
% Figure 4
\begin{figure}[!htp]
\label{fig:fig4}
\centering
\vspace{-3mm}
\includegraphics[width=3.6in]{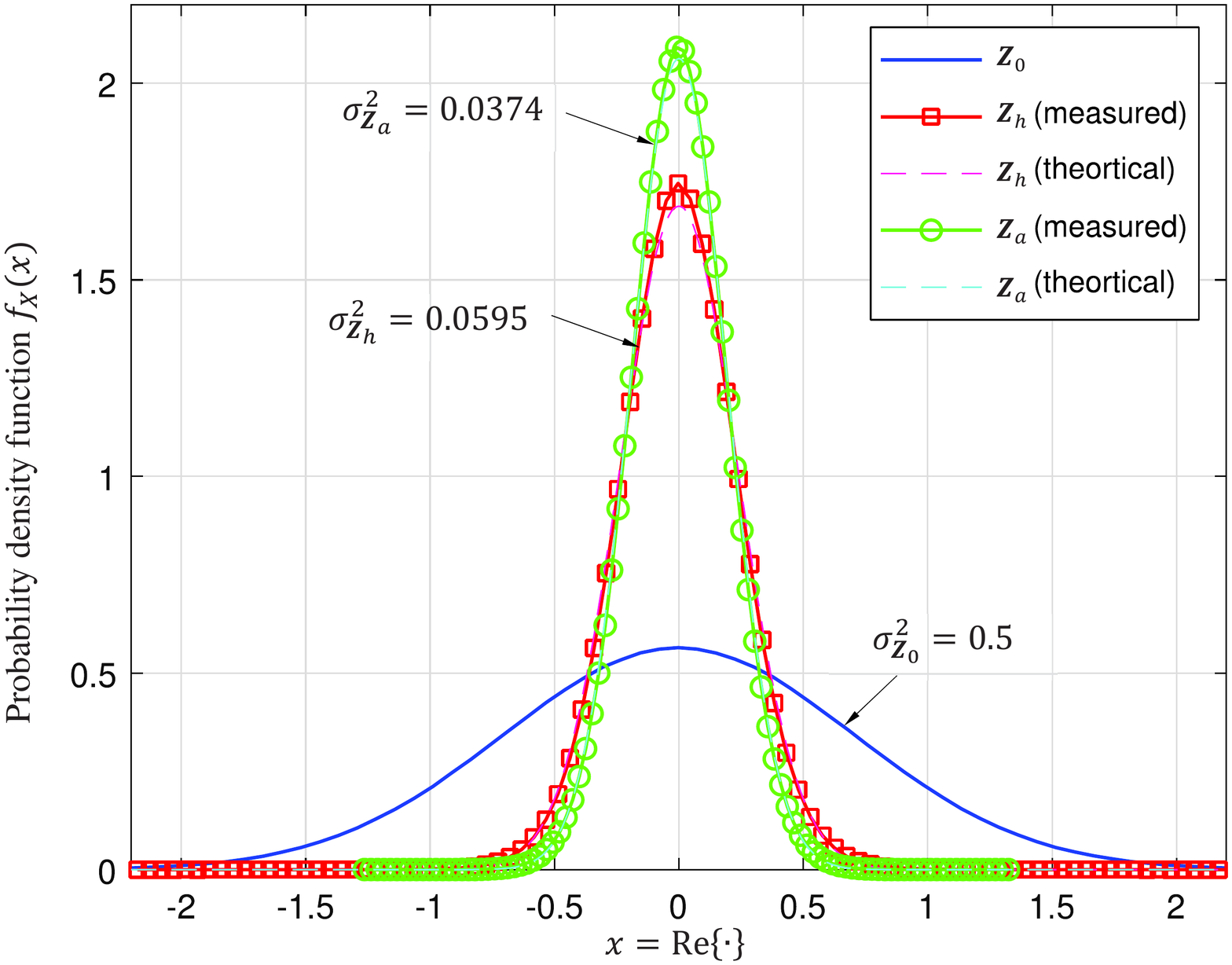}
\vspace{-7mm}
\caption{Probability density function of post-processing noise after \emph{Subspace Training} and combining at 0dB SNR}
\vspace{-5mm}
\end{figure}This property can be illustrated in Fig.~4. Let us study the Probability Density Function (PDF) of the real part signal of the raw AWGN $\mathbf{Z}_0$, as well as that of the channel estimation error after horizontal \emph{Subspace Training} $\mathbf{Z}_h$ and the channel estimation error after arithmetic mean $\mathbf{Z}_a$. Notice that we can plot the PDFs straightforwardly by means of measuring $\mathbf{Z}_h$ and $\mathbf{Z}_a$, or alternatively, plot the theoretical Gaussian distribution functions by the measured post-processing noise variance. The PDFs, obtained from both approaches, coincides to each other very well, which confirms that the channel estimation errors still obey Gaussian distribution with reduced variance. Let these noisy channel estimates replace old observations of the training data, and produce the new channel estimates repeatedly with the same NNs by means of \emph{Subspace Training}. This iterative procedure is referred to as \emph{Turbo-AI}, due to the iteratively achievable processing gain, which eventually makes the channel estimates get very close to the solution produced by \emph{genie-aided} channel estimator. 

Note that the \emph{Turbo-AI} procedure introduces a new way to train the NN. As opposed to training the whole layers (from all iterations) at once, we train each layer to the same final labels, from bottom to top. This acts like a regularization by limiting the degrees of freedom that our network can take. It further boosts the convergence speed. Moreover, the chance of getting stuck in a local minima is a lot smaller this way, as in each stage we are dealing with a small 2-layered NN. This solution is not entirely heuristic since it is based on the observation that the dominant randomness at the input and the output of \emph{Turbo-AI} is Gaussian with different variances.

In Fig.~5, the training phase for $i$-th iteration of \emph{Turbo-AI} is presented. Notice that a dedicated set of NN weights, so-called NN-models, can be stored for the validation from other users with the same spatial characteristics. In the inference phase, depicted in Fig.~6, the user should estimate the initial SNR, decide to start with the best matched pre-trained NN-models $\mathbf{W}_{v,i}$ and $\mathbf{W}_{h,i}$, and replace both \emph{Subspace Training} blocks in Fig.~5 to estimate the channel iteratively. Considering the potential complexity for hardware design, introduced by the square root operations to calculate geometric mean for both $\mathbf{W}_{v,i}$ and $\mathbf{W}_{h,i}$ during the iterations, we adopt arithmetic mean  for subspace combining to realize \emph{Turbo-AI} as an overall relatively low complex solution throughout this paper. In \emph{Appendix B}, the proof of the applicability of \emph{Turbo-AI} will be provided. 
% Figure 5
\begin{figure}[!htp]
\label{fig:fig5}
\centering
\vspace{-8mm}
\includegraphics[width=3.5in]{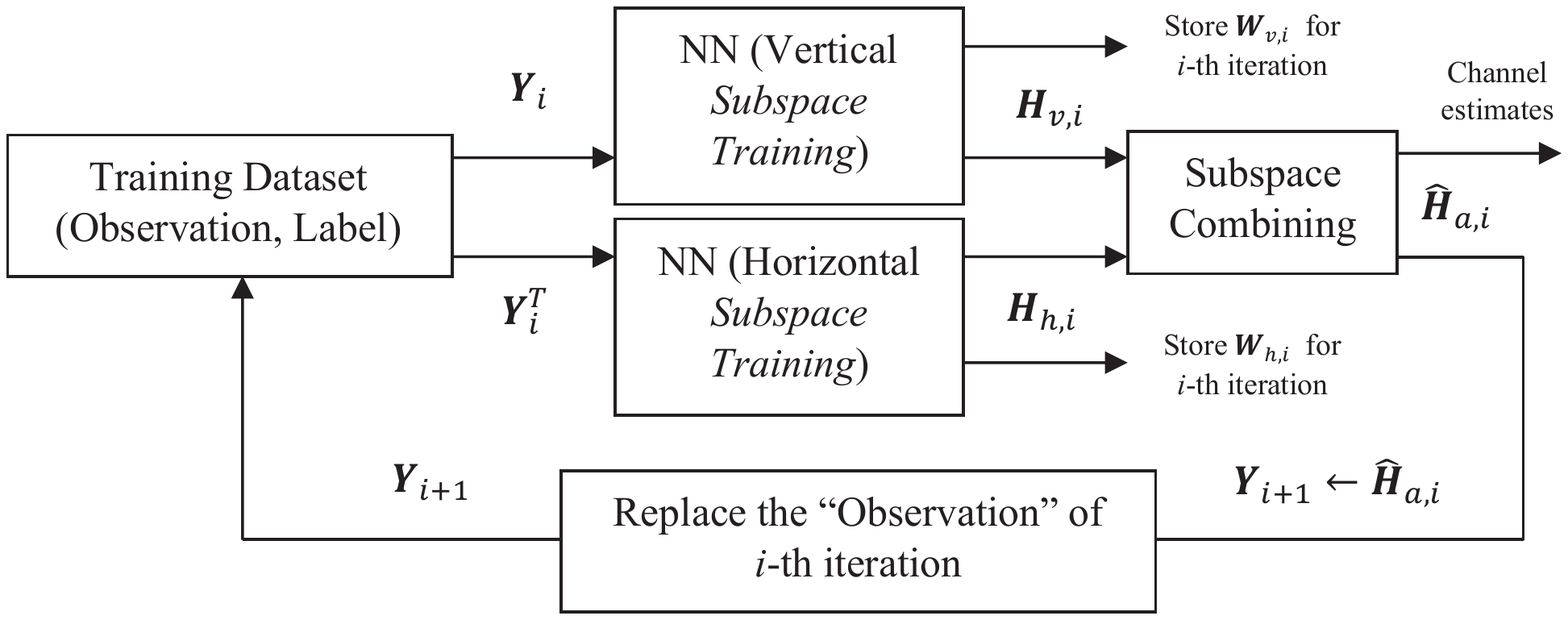}
\vspace{-91mm}
\caption{Training phase: Block diagram for $i$-th iteration training of \emph{Turbo-AI}}
\vspace{-1mm}
\end{figure}
% Figure 6
\begin{figure}[!htp]
\label{fig:fig6}
\centering
\vspace{-7mm}
\includegraphics[width=3.5in]{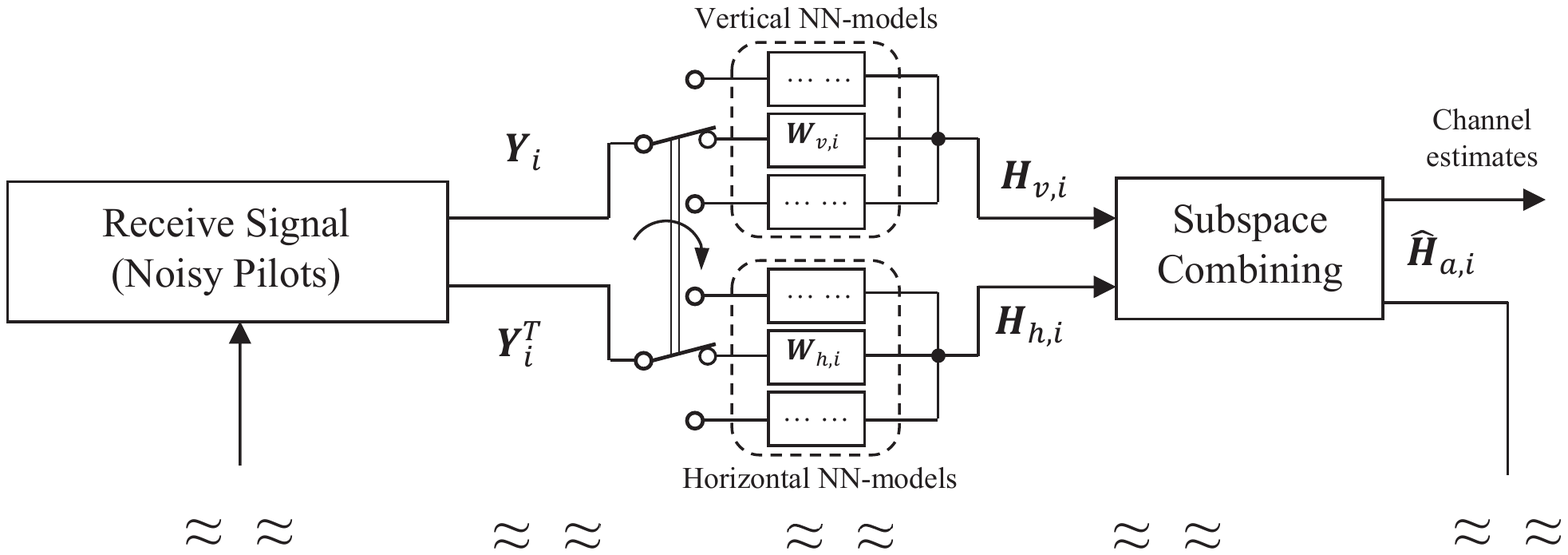}
\vspace{-95mm}
\caption{Inference phase: Block diagram for $i$-th iteration training of \emph{Turbo-AI}}
\vspace{-1mm}
\end{figure}

\subsection{Universal Training for Turbo-AI}
\emph{Turbo-AI} can iteratively improve the channel estimation. The training phase, depicted in Fig.~5, executes dedicated training for one direction with respect to vertical and horizontal spatial domain. For another direction, the vertical and horizontal NNs have to be retrained again. For the same reason, a set of vertical and horizontal NN-models have to be stored, because the effective SNR will increase during the iterations. Especially, the estimation of initial SNR to start the \emph{Turbo-AI} in Fig.~6 should be precise. Otherwise, it can also cause the mismatch of the NN-models during the iterations. This motivates us to think about the question, whether it is possible to span the parameter space of the training, and make the NN-models of \emph{Turbo-AI} become more universal and more robust against the parameter estimation error in the practice. Thus, we name this concept \emph{Universal Training}. In Fig.~7, 
% Figure 7
\begin{figure}[!htp]
\label{fig:fig7}
\centering
\vspace{-2mm}
\includegraphics[width=3.5in]{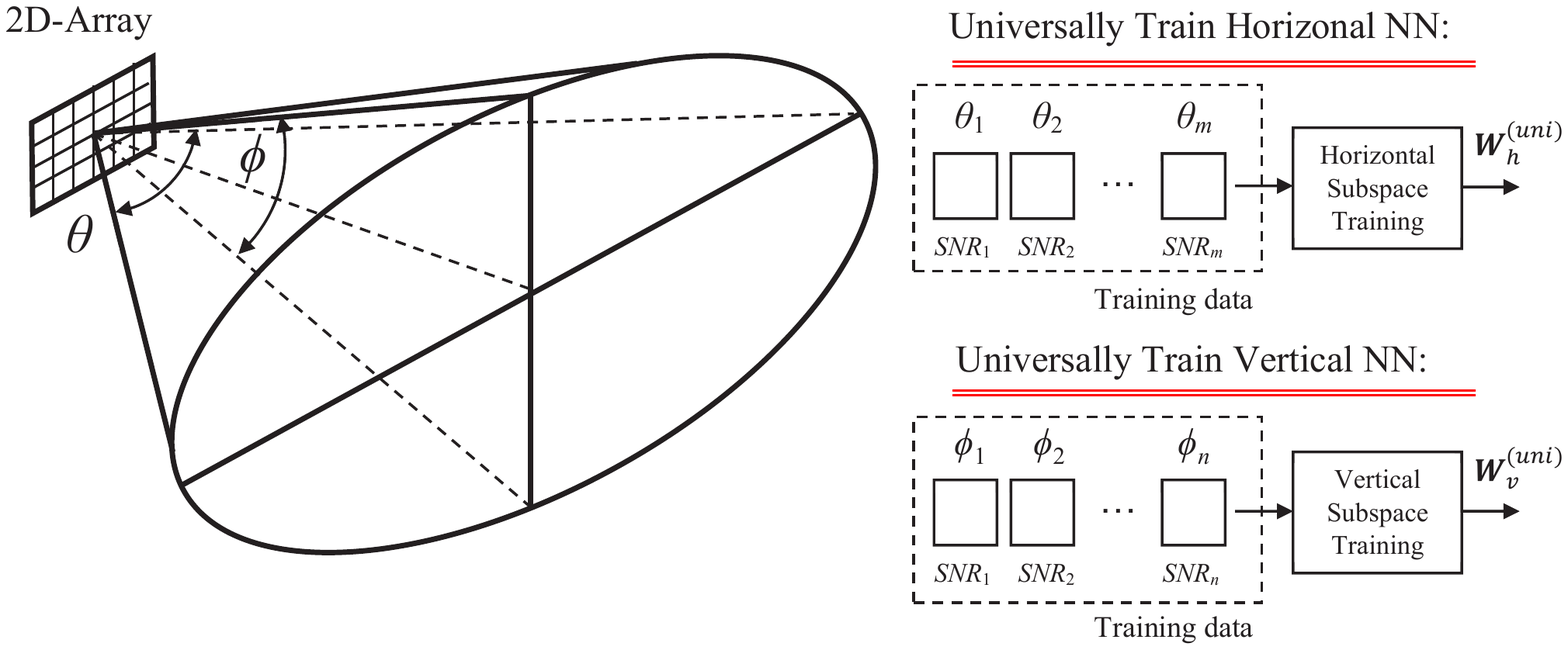}
\vspace{-94mm}
\caption{\emph{Universal Training} with respect to SNR and spatial domain}
\vspace{-2mm}
\end{figure}the basic idea of \emph{Universal Training} is illustrated. For training the horizontal NN, the training data consists of the data batches for different horizontal angle $\theta_m$ and SNR values randomly. During the training stage, the NN tries to find out a model to deliver sub-optimum solution for all possible combination of parameters $\theta_m$ and $\textrm{SNR}_m$. The randomization for the data combination is necessary to prevent ML from falling in a local optimum. Thus, the learning procedure takes longer than the conventional training. 

After the training, a universal NN-model is ready for channel estimation within the parameter space spanned by data combination $\theta_m$ and $\textrm{SNR}_m$. Similarly, the vertical NN-model can be universally learned in another parameter space, spanned by parameter $\phi_n$ and $\textrm{SNR}_n$. As a matter of fact, the \emph{Universal Training} is sub-optimal and can cause performance loss, if more parameters are introduced in training. On the other side, with \emph{Universal Training}, \emph{Turbo-AI} requires only a couple of vertical NN-models and horizontal NN-modes, or even one vertical NN-model and one horizontal NN-model to enable the iterations for arbitrary combination of $\theta_m$, $\phi_n$ and SNR value. This can introduce huge cost saving and flexibility in hardware design for a practical implementation. 

\section{Simulations and Numerical Results}
In this section, let us focus on the performance of \emph{Turbo-AI} with link level simulations. Throughout our investigation, the spatial i.i.d. Rayleigh fading channel, modeled by (4) and (5), is adopted to generate channel responses. In Table~I, the parameters are summarized. 
% Table I
\begin{table}[!htp]
\vspace{-1mm}
%% increase table row spacing, adjust to taste
\renewcommand{\arraystretch}{1.1}
% if using array.sty, it might be a good idea to tweak the value of
% \extrarowheight as needed to properly center the text within the cells
\caption{Link Level Simulation And Training Parameters}
\vspace{-3mm}
%\label{table_example}
\centering
%% Some packages, such as MDW tools, offer better commands for making tables
%% than the plain LaTeX2e tabular which is used here.
\begin{tabular}{|c||c|}
\hline
\hline
Number of vertical elements & $M = 8$ \\
\hline
Number of horizontal elements & $N = 16$ \\
\hline
Antenna spacing & Half of wavelength \\
\hline
Vertical Angular spread & $\pi/180$ \\
\hline
Horizontal Angular spread & $\pi/90$ \\
\hline
Optimizer for ML & Adam\\
\hline
Learning rate & $\alpha = 0.009$\\
\hline
Hyper-parameters &  $\beta_1 = 0.9$, $\beta_2 = 0.999$\\
\hline
Activation function & ReLu \\
\hline
Training data set size & $K = 500000$ (Time domain)\\
\hline
\hline
\end{tabular}
\vspace{-1mm}
\end{table}

% Figure 8
\begin{figure}[!htp]
\label{fig:fig8}
\centering
\vspace{-0mm}
\includegraphics[width=3.6in]{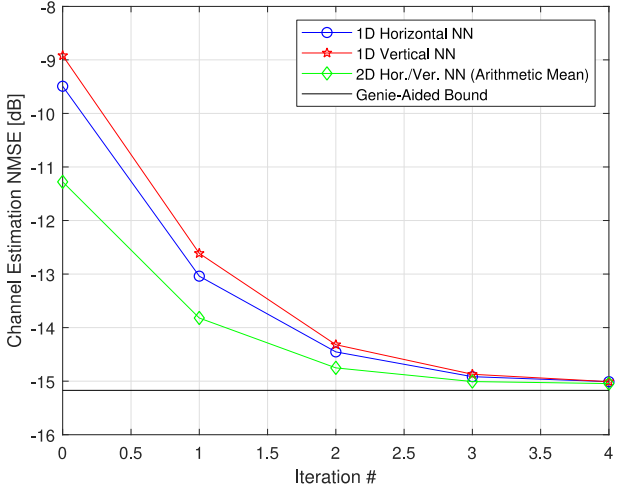}
\vspace{-8mm}
\caption{\emph{Turbo-AI} performance: channel estimation NMSE versus iterations, a 2D-array with $M = 8$ and $N=16$, i.i.d. Rayleigh fading, SNR at 0dB}
\vspace{-2mm}
\end{figure}In Fig.~8, the performance of \emph{Turbo-AI} is presented for user validation with the stored NN-models through \emph{Subspace Training}, as depicted in Fig.~6. With the cooperation between vertical and horizontal NN, the arithmetic mean based subspace combining can approach the \emph{genie-aided} MMSE estimator very fast, especially at the first and the second iteration. Even at relatively low SNR, e.g. SNR at 0dB, the performance gap to the \emph{genie-aided} bound is only 0.13dB after four iterations. In Fig.~9, the measured PDFs of the channel estimates from horizontal domain after the iterations are presented, which are a set of Gaussian distribution curves with reduced variance. 
% Figure 9
\begin{figure}[!htp]
\label{fig:fig9}
\centering
\vspace{-0mm}
\includegraphics[width=3.7in]{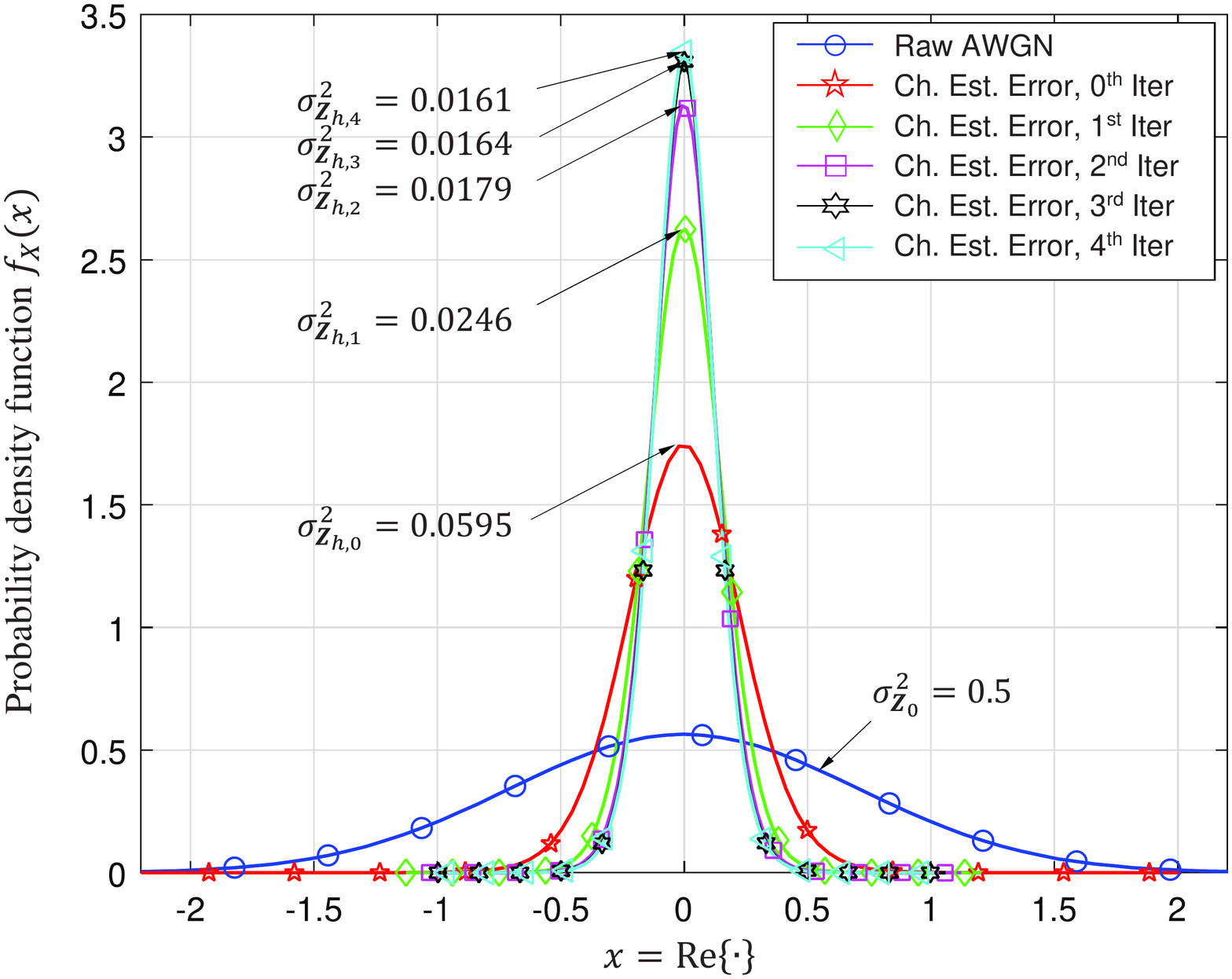}
\vspace{-8mm}
\caption{\emph{Turbo-AI} performance: track the PDFs of horizontal estimates, a 2D-array with $M = 8$ and $N=16$, i.i.d. Rayleigh fading, SNR at 0dB}
\vspace{-6mm}
\end{figure}

In Fig.~10, the \emph{Turbo-AI} performance with \emph{Universal Training} is summarized. As mentioned in previous section, the more parameters are introduced in the training, the more performance loss should be expected. In Fig.~10, we consider three parameters, namely the range of SNR, horizontal and vertical spatial coverage $\theta$ and $\phi$, respectively. First of all, let us carry out the \emph{Universal Training} only for SNR with the range from 0dB to 15dB, and assume that $\theta$ and $\phi$ are \emph{a priori} known directions. Comparing to the dedicated training, the performance loss uniquely introduced by the universal SNR training is about 1.20dB. 
% Figure 10
\begin{figure}[!htp]
\label{fig:fig10}
\centering
\vspace{0mm}
\includegraphics[width=3.6in]{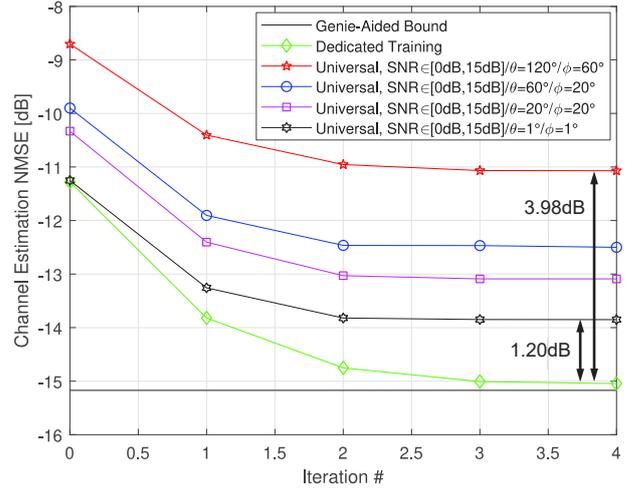}
\vspace{-8mm}
\caption{\emph{Turbo-AI} performance: dedicated training versus universal training, a 2D-array with $M = 8$ and $N=16$, i.i.d. Rayleigh fading, SNR at 0dB}
\vspace{-5mm}
\end{figure}The universal SNR training is quite meaningful for \emph{Turbo-AI}, since only one NN-model is required for the \emph{Turbo-AI}, due to being able to adapt to all possible effective SNR values during the iterations. Then, let us increase the coverage horizontally and vertically with respect to both parameters $\theta$ and $\phi$. We can observe further obvious degradation. Finally, we carry out \emph{Universal Training} for the whole sector, namely $-60^\textrm{o}<\theta_m<60^\textrm{o}$, $30^\textrm{o}<\phi_n<90^\textrm{o}$ and $\textrm{SNR} \in [0\textrm{dB},15\textrm{dB}]$. It is quite convenient to exploit \emph{Turbo-AI} to estimate the channel with one vertical NN-model and one horizontal NN-model for the whole sector. Being the trade-off between complexity and performance, the performance gap to dedicated training is about 3.98dB. Nevertheless, comparing to the Least Square (LS) channel estimator, which achieves only 0dB channel estimation NMSE at 0dB SNR, \emph{Turbo-AI} with one NN-model converges at approximately -11dB, which is quite satisfactory. 
% Figure 11
\begin{figure}[!htp]
\label{fig:fig11}
\centering
\vspace{-2mm}
\includegraphics[width=3.6in]{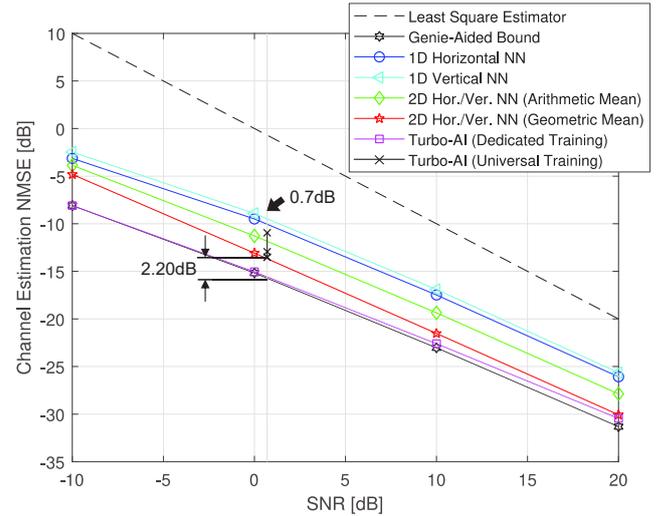}
\vspace{-9mm}
\caption{Channel estimation NMSE versus SNR, a 2D-array with $M = 8$ and $N=16$, i.i.d. Rayleigh fading}
\vspace{-2mm}
\end{figure}
In Fig.~11, the performance of diverse channel estimators is summarized. The ML-based channel estimators are bounded by LS estimator and \emph{genie-aided} MMSE estimator. The 2D \emph{Subspace Training} based channel estimators outperform the horizontal or vertical NNs. Especially, the 2D \emph{Subspace Training} and combining with geometric mean outperforms that with arithmetic mean, which numerically verifies equation (15) to (17) in the previous section. \emph{Turbo-AI} with dedicated training exhibits powerful performance. With four iterations, the \emph{genie-aided} bound can be tightly approached, especially for low SNR region. Further, with the training parameters $10^\textrm{o}<\theta_m<30^\textrm{o}$, $40^\textrm{o}<\phi_n<60^\textrm{o}$ and $\textrm{SNR} \in [0\textrm{dB},15\textrm{dB}]$, we verify the pre-trained universal NN-model for a SNR value at 0.7dB, a strange value, which is not included in our selected universal training data set at all. Simulation exhibits that this universal NN-model for \emph{Turbo-AI} is still applicable. With three iterations, the performance gap to \emph{genie-aided} bound is only 2.20dB. 
\vspace{1mm}
\section{Conclusion}
In this paper, we propose a set of ML-based channel estimation approaches for 2D massive arrays. Exploiting the 2D Kronecker channel covariance model, we can carry out vertical and horizontal \emph{Subspace Training}, to estimate the channel in sense of arithmetic mean and geometric mean. \emph{Subspace Training} provides a sub-optimal solution with much lower complexity than that of full array size inputs. Although the cost saving from \emph{Subspace Training} with respect to hardware implementation is huge,  a performance gap does still exist, considering the \emph{genie-aided} MMSE channel estimator. Thus, we further trade off the complexity and performance, and introduce an iterative ML-based channel estimation approach, referred to as \emph{Turbo-AI}, in which the observation for ML can be iteratively de-noised by means of the cooperation of horizontal and vertical \emph{Subspace Training} and combining. Numerical results exhibit that the \emph{genie-aided} bound performance can be tightly approached by \emph{Turbo-AI}, even at very low SNR. From the view point of system design, \emph{Subspace Training} and \emph{Turbo-AI} provide diverse application possibilities to support both offline learning and online learning with scalable complexity. Furthermore, in order to enhance the flexibility of hardware design, \emph{Universal Training} is introduced as a counterpart for training with dedicated parameters. In \emph{Universal Training}, we span the parameter space in training phase and make the NN-model adapt to a wide range of parameters. For instance, the parameter space can be spanned by vertical spatial coverage $\phi$, horizontal spatial coverage $\theta$ and SNR. It is demonstrated that flexible performance can be achieved by adjusting the range of the parameter space. Thus, \emph{Universal Training} provides additional degrees of freedom to facilitate practical implementation. For the future perspective, we will validate the concepts, e.g. \emph{Subspace Training}, \emph{Turbo-AI} and \emph{Universal Training}, in 3GPP spatial channel model for a multicarrier system. The channel correlation in space, time and frequency can be jointly considered. These additional dimensions are assumed to be able to provide further enhancement for ML-based channel estimation. 

\vspace{4mm}
\section*{Appendix}
\subsection{Proof of (15)}
Before proving equation (15), let us introduce several \emph{Lemmas} to define the NN, which is exploited for channel estimation. We inherit the notations from previous sections. The observations and labels of a 2D-array with $M$ vertical elements and $N$ horizontal elements are denoted as $M \times N$ matrices $\mathbf{Y}$ and $\mathbf{H}$, respectively, with $\mathbf{H}_v = \mathbf{H}$ and $\mathbf{H}_h = \mathbf{H}^T$. Let $\mathbf{Z}$ denote the $M \times N$ AWGN matrix, with covariance matrix $\mathbf{R}_Z = N_0\mathbf{I}_M = 2\sigma^2\mathbf{I}_M$. Without loss of generality, we adopt the vertical subspace for defining the \emph{Lemmas}. The same \emph{Lemmas} hold for horizontal domain as well.  
\vspace{4mm}
\begin{lemma}When a neural network is trained over adequate amount of channel observation, sampled from one given distribution, this neural network is called converged, if following properties hold,
% Equation 21
 \vspace{2mm}
\begin{equation}
E [\|\mathbf{W}_v \mathbf{Y} - \mathbf{H}\|_F^2] \leq E \big[ \|\mathbf{Z}\|_F^2 \big]
\vspace{-2mm}
\end{equation}
% Equation 22
 \vspace{-4mm}
\begin{equation}
\|\mathbf{W}_v - \mathbf{W}_{v,\textrm{genie}}\|_F < \epsilon
\vspace{2mm}
\end{equation}
where $\mathbf{W}_v$ denotes the $M \times M$ weighting matrix and $\mathbf{W}_v \mathbf{Y} - \mathbf{H}$ represents the zero-mean estimation error. The operator $\|\mathbf{A}\|_F$ denotes the Frobenius norm of matrix $\mathbf{A}$. Matrix $\mathbf{W}_{v,\textrm{genie}}$ denotes the optimal MMSE weighting matrix in vertical subspace, defined in (6), and $\epsilon$ is small non-negative value. \emph{Lemma 1} follows from the the MSE loss function. 
\end{lemma}

 \vspace{4mm}
Equation (22) in \emph{Lemma 1} is difficult to be proven mathematically, as the proof of convergence to the optimum point generally does not exist for an arbitrary architecture and loss function. The conditions in \emph{Lemma 1} are based on our observations of training the NNs on many different scenarios and channel data distributions, corrupted by Gaussian noise. We do not claim these conditions and properties for an arbitrary type of data. Nevertheless, the procedure for the core NN design based on \cite{ref7.00} and our simulations
indicates that it holds for Gaussian inputs. Furthermore, we treat $\mathbf{W}_v$ as a deterministic matrix in proving (15), if the NN is converged and the matrix $\mathbf{W}_{v,\textrm{genie}}$ itself is deterministic as well. 

\vspace{4mm}
\begin{lemma}
From \emph{Lemma 1}, it straightforwardly holds
% Equation 22a
\vspace{1mm}
\begin{equation}
E [\|\mathbf{W}_v \mathbf{Z}\|_F^2] \leq E \big[ \|\mathbf{Z}\|_F^2 \big]\textrm{.} \nonumber
\vspace{1mm}
\end{equation}
Furthermore, for an arbitrary vertical array element $i$, with $i \in \{1,2,\cdots,M\}$, through the spatial filtering with $\mathbf{W}_v$, the post-processing noise will be reduced, if the convergence of NN is reached, namely 
% Equation 22b
\vspace{1mm}
\begin{equation}
\label{eq22}
E [\|\mathbf{W}_v^{(i,:)} \mathbf{Z}\|_F^2] \leq E \big[ \|\mathbf{Z}^{(i,:)}\|_F^2 \big]\textrm{,} 
\vspace{1mm}
\end{equation}where $\mathbf{W}_v^{(i,:)}$ and $\mathbf{Z}^{(i,:)}$ denote the $i$-th row of matrices $\mathbf{W}_v$ and $\mathbf{Z}$, respectively. 
\end{lemma}
\vspace{4mm}
With the \emph{Lemma 1} and \emph{Lemma 2} introduced above, now we can start to prove equation (15). 
\vspace{4mm}
\begin{proof}
With \emph{Lemma 2}, we obtain
% Equation 23a
\vspace{2mm}
\begin{equation}
\label{eq23a}
E \big[\|\mathbf{W}_v \mathbf{Z}\|_F^2\big]=E \big[ \textrm{Tr}(\mathbf{W}_v\mathbf{Z}\mathbf{Z}^H\mathbf{W}_v^H)\big]=2\sigma^2 \textrm{Tr}(\mathbf{W}_v\mathbf{W}_v^H) \nonumber
\vspace{-2mm}
\end{equation}
% Equation 23b
\vspace{-2mm}
\begin{equation}
\label{eq23b}
\leq E \big[ \|\mathbf{Z}\|_F^2 \big] = E \big[ \textrm{Tr}(\mathbf{Z}\mathbf{Z}^H)\big]=2\sigma^2M\textrm{.}
\vspace{2mm}
\end{equation}Thus, it yields,
% Equation 24
\vspace{1mm}
\begin{equation}
\label{eq24}
\textrm{Tr}(\mathbf{W}_v\mathbf{W}_v^H) \leq M \Leftrightarrow \sum_{i=1}^M \sum_{m=1}^M |\mathbf{W}_v^{(i,m)}|^2 \leq M \textrm{,}
\vspace{0mm}
\end{equation}where $\mathbf{W}_v^{(i,m)}$ denotes the weighting coefficient on $i$-th row and $m$-th column of $\mathbf{W}_v$. Exploiting the same operations for (23), considering (24) and (25), we obtain,
% Equation 25
\vspace{1mm}
\begin{equation}
\label{eq25}
0 \leq \sum_{m=1}^M |\mathbf{W}_v^{(i,m)}|^2 \leq 1 \textrm{, }\forall i \in \{1,2,\cdots,M\} \textrm{.}
\vspace{0mm}
\end{equation}Similarly, for horizontal case, we have
% Equation 26
\vspace{1mm}
\begin{equation}
\label{eq26}
0 \leq \sum_{n=1}^N |\mathbf{W}_h^{(j,n)}|^2 \leq 1 \textrm{, }\forall j \in \{1,2,\cdots,N\} \textrm{.}
\vspace{1mm}
\end{equation}
Considering arithmetic mean to combine the estimates from vertical and horizontal subspaces, we have
% Equation 27
\vspace{1mm}
\begin{equation}
\label{eq27}
\hat{\mathbf{H}}_a^{(i,j)} = \frac{1}{2} \big[ \hat{\mathbf{H}}_v^{(i,j)} + \hat{\mathbf{H}}_h^{(j,i)}\big] \textrm{.}
\vspace{1mm}
\end{equation}Furthermore, it holds,
% Equation 28a
\vspace{1mm}
\begin{equation}
\label{eq28a}
E\big[|\hat{\mathbf{H}}_a^{(i,j)}|^2\big] =\frac{1}{4} E\big[ | \hat{\mathbf{H}}_v^{(i,j)} + \hat{\mathbf{H}}_h^{(j,i)}|^2\big] \nonumber
\vspace{-2mm}
\end{equation}
% Equation 28b
\vspace{-2mm}
\begin{equation}
\label{eq28b}
=\underbrace{\frac{1}{4}E\big[|\hat{\mathbf{H}}_v^{(i,j)}|^2\big]}_{A} +\underbrace{\frac{1}{4}E\big[|\hat{\mathbf{H}}_h^{(j,i)}|^2\big]}_{B} + \underbrace{\frac{1}{2}\textrm{Re}\big\{  E\big[ \hat{\mathbf{H}}_v^{(i,j)} \hat{\mathbf{H}}_h^{(j,i)^*}\big]\big\}}_C \textrm{.}
\vspace{1mm}
\end{equation}Now, let us resolve the terms $A$, $B$ and $C$ in the equation above. It holds,
% Equation 29 to 31
\vspace{1mm}
\begin{align}
\label{eq29}
A =& \frac{1}{4}E\bigg[\big| \sum_{m=1}^M \mathbf{W}_v^{(i,m)} \mathbf{H}_v^{(m,j)} \big|^2\bigg] + \frac{\sigma^2}{2} \sum_{m=1}^M \big| \mathbf{W}_v^{(i,m)} \big|^2 \\
B =& \frac{1}{4}E\bigg[\big| \sum_{n=1}^N \mathbf{W}_h^{(j,n)} \mathbf{H}_h^{(n,i)} \big|^2\bigg] + \frac{\sigma^2}{2} \sum_{n=1}^N \big| \mathbf{W}_h^{(j,n)} \big|^2 \\
C =& \frac{1}{2} \textrm{Re} \bigg\{ E\bigg[ \sum_{m=1}^M \mathbf{W}_v^{(j,m)} \mathbf{H}_v^{(m,i)} \sum_{n=1}^N \mathbf{W}_h^{(j,n)^*} \mathbf{H}_h^{(n,i)^*} \bigg] \bigg\} \nonumber \\
& + \sigma^2 \mathbf{W}_v^{(i,i)}\mathbf{W}_h^{(j,j)} \textrm{.}
\end{align}
\vspace{1mm}
Let us pick out the post-processing AWGN related terms from $A$, $B$ and $C$. It holds,
% Equation 32
\vspace{2mm}
\begin{align}
\label{eq32}
\textrm{VAR}\big[\mathbf{Z}_a^{(i,j)} \big] =& \frac{\sigma^2}{2} \sum_{m=1}^M |\mathbf{W}_v^{(i,m)} |^2+\frac{\sigma^2}{2} \sum_{n=1}^N |\mathbf{W}_h^{(j,n)} |^2   \nonumber \\
& + \sigma^2 \mathbf{W}_v^{(i,i)}\mathbf{W}_h^{(j,j)} \textrm{,}
\vspace{2mm}
\end{align}where the $M \times N$ matrix $\mathbf{Z}_a$ denotes the post-processing noise after deploying arithmetic mean. Consider the following inequality
% Equation 33a
\vspace{1mm}
\begin{equation}
\label{eq33a}
\sum_{m=1}^M |\mathbf{W}_v^{(i,m)} |^2+\sum_{n=1}^N |\mathbf{W}_h^{(j,n)} |^2  \nonumber
\vspace{-1mm}
\end{equation}
% Equation 33b
\vspace{-1mm}
\begin{equation}
\label{eq33b}
\geq \mathbf{W}_v^{(i,i)^2}+\mathbf{W}_h^{(j,j)^2}\geq 2\mathbf{W}_v^{(i,i)}\mathbf{W}_h^{(j,j)}\textrm{.}
\vspace{1mm}
\end{equation}With (33) and (34), the upper bound of $\textrm{VAR}\big[\mathbf{Z}_a^{(i,j)} \big]$ can be represented as
% Equation 34
\vspace{2mm}
\begin{align}
\label{eq34}
\textrm{VAR}\big[\mathbf{Z}_a^{(i,j)} \big] & \leq 2\big\{\sum_{m=1}^M |\mathbf{W}_v^{(i,m)} |^2+\sum_{n=1}^N |\mathbf{W}_h^{(j,n)} |^2  \big\} \frac{\sigma^2}{2} \nonumber \\
 & \leq 2\sigma^2 =N_0 \textrm{.}
\vspace{2mm}
\end{align}With (33), we find the lower bound of $\textrm{VAR}\big[\mathbf{Z}_a^{(i,j)} \big]$ as
% Equation 35
\vspace{2mm}
\begin{align}
\label{eq35}
\textrm{VAR}\big[\mathbf{Z}_a^{(i,j)} \big] & > \frac{\sigma^2}{2} \sum_{m=1}^M |\mathbf{W}_v^{(i,m)} |^2+\frac{\sigma^2}{2} \sum_{n=1}^N |\mathbf{W}_h^{(j,n)} |^2 \nonumber \\
 & = \frac{\sigma^2}{2}\rho =\frac{\rho}{4}N_0 \textrm{.}
\vspace{2mm}
\end{align}with $\rho=\sum_{m=1}^{M} |\mathbf{W}_v^{(i,m)}|^2+\sum_{n=1}^{N} |\mathbf{W}_h^{(j,n)}|^2$ and $0 \leq \rho \leq 2$. Finally, let us make a summary, by jointly considering (35) and (36). It holds,
% Equation 36
\vspace{2mm}
\begin{equation}
\label{eq36}
\frac{\rho}{4}N_0 < \textrm{VAR}\big[\mathbf{Z}_a^{(i,j)} \big] \leq N_0 \textrm{.}
\vspace{2mm}
\end{equation}
(q.e.d.)
\end{proof}                
\vspace{0mm}
\subsection*{Further Discussion}
In Fig.~12, numerical demonstration is provided to exhibit the contour of $|\mathbf{W}_v^{(i,j)}|^2$ with $M=8$. The presented weight matrix is the average of a large number of weight matrices, generated by the pre-trained NN-model for certain stationary statistics. Thus, the weight matrices in Fig.~12 can be regarded as the approximation of the optimum $|\mathbf{W}_{v,\textrm{genie}}^{(i,j)}|^2$ within different batch stationary inference stages. In i.i.d. data batch, no spatial correlation can be exploited, and in the data batch involving Kronecker model, rich spatial correlation exists. Fig.~12 not only verifies the equations (25)-(27) numerically, but also reveals the fact, that the off-diagonal elements in $\mathbf{W}_v$ and  $\mathbf{W}_h$ will become dominant, if the spatial correlation increases. This makes the term $\mathbf{W}_v^{(i,i)}\mathbf{W}_h^{(j,j)}$ in (33) reduce and become negligible, and helps to reduce the post-processing noise variance and improve the quality of channel estimates. Let us again focus on post-processing noise variance in (33), and introduce following notations for further simplification as
% Equation 37 to 40
\vspace{-2mm}
\begin{align}
z & = \frac{1}{\sigma^2} \textrm{VAR}\big[\mathbf{Z}_a^{(i,j)} \big] \textrm{,} \\
x & = \sum_{m=1}^M  \big| \mathbf{W}_v^{(i,m)} \big|^2  \textrm{,} \\
y & = \sum_{n=1}^N  \big| \mathbf{W}_h^{(j,n)} \big|^2  \textrm{,} \\
d & =  \mathbf{W}_v^{(i,i)}\mathbf{W}_h^{(j,j)}  \textrm{.}
\end{align}
\vspace{0mm}Obviously, equation (33) can be thus represented as
% Equation 41
\vspace{0mm}
\begin{equation}
\label{eq41}
z =  \frac{1}{2}(x+y)+d\textrm{.}
\vspace{0mm}
\end{equation}Furthermore, considering the post-processing noise variance $\textrm{VAR}\big[ \mathbf{Z}_v^{(i,j)} \big]$ and $\textrm{VAR}\big[ \mathbf{Z}_h^{(i,j)} \big]$, which are uniquely obtained from vertical and horizontal \emph{Subspace Training}, respectively. With equation (29) to equation (31), let us additionally introduce, 
% Figure 2.5
\begin{figure}[!htp]
\vspace{-3mm}
\centering
\includegraphics[width=3.6in]{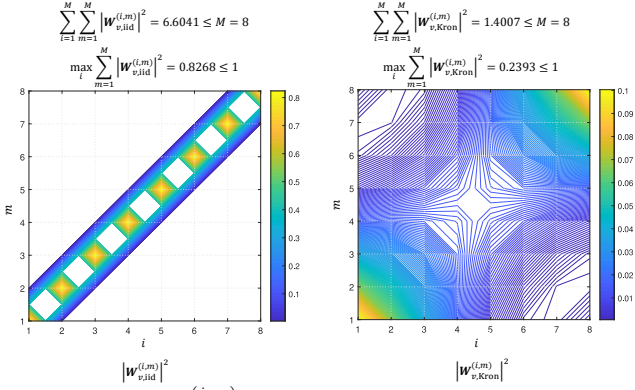}
\vspace{-23mm}
\caption{Contour of $|\mathbf{W}_v^{(i,m)}|^2$ with and without spatial correlation at 10dB SNR}
\vspace{-4mm}
\end{figure}
% Equation 42 to 43
\vspace{0mm}
\begin{align}
\label{eq42}
\frac{1}{\sigma^2}\textrm{VAR}\big[ \mathbf{Z}_v^{(i,j)} \big] & = 2 \sum_{i=0}^M \big| \mathbf{W}_v^{(i,m)} \big|^2 = 2x\textrm{,} \\
\frac{1}{\sigma^2}\textrm{VAR}\big[ \mathbf{Z}_h^{(i,j)} \big] & = 2 \sum_{j=0}^N \big| \mathbf{W}_h^{(j,n)} \big|^2 = 2y\textrm{.}
\vspace{0mm}
\end{align}
Thus, we are able to find the effective subspace combining region for $(x,y)$-pairs, which realizes both inequalities $\textrm{VAR}\big[ \mathbf{Z}_a^{(i,j)}\big] < \textrm{VAR}\big[ \mathbf{Z}_v^{(i,j)}\big]$ and $\textrm{VAR}\big[ \mathbf{Z}_a^{(i,j)}\big] < \textrm{VAR}\big[ \mathbf{Z}_h^{(i,j)}\big]$ for the post-processing noise variance in vertical and horizontal domains individually, so that the boosting effect in Fig.~8 by means of subspace combining can be achieved, if \emph{Turbo-AI} is additionally deployed. With the introduced simplified notations, following equation system can be established as,
% Equation 44
\vspace{4mm}
\begin{equation}
\label{eq44}
\left\{
\begin{aligned}
z & = \frac{1}{2}(x+y)+d \textrm{,}\\
z & < 2x \textrm{,} \\
z & < 2y \textrm{,}\\
\end{aligned}
\right.
\vspace{2mm}
\end{equation}with $0 \leq x \leq 1$, $0 \leq y \leq 1$ and $0 \leq d \leq 1$.  In Fig.~13, the equation system (45) is sketched in Cartesian coordinates. The blue intersection part denotes the $(x,y,z)$-pairs, that satisfy the equation (45), and its corresponding projection on plane $z=d$ is illustrated as the red shadow. With Fig.~13, we obtain following inequality 
 % Equation 45a
\vspace{1mm}
\begin{equation}
\label{eq45a}
\frac{1}{3}\sum_{m=1}^M  \big| \mathbf{W}_v^{(i,m)} \big|^2 +\frac{2}{3}  \mathbf{W}_v^{(i,i)}\mathbf{W}_h^{(j,j)} < \sum_{n=1}^N \big| \mathbf{W}_h^{(j,n)} \big|^2 \nonumber
\vspace{-2mm}
\end{equation}
 % Equation 45b
\begin{equation}
\vspace{-2mm}
\label{eq45b}
< 3\sum_{m=1}^M \big| \mathbf{W}_v^{(i,m)} \big|^2 -2 \mathbf{W}_v^{(i,i)}\mathbf{W}_h^{(j,j)} \textrm{.}
\vspace{2mm}
\end{equation}Notice that the equation (46) provides the range for both $\sum_{m=1}^M | \mathbf{W}_v^{(i,m)} |^2$ and $\sum_{n=1}^N | \mathbf{W}_h^{(j,n)} |^2$, in order to effectively support \emph{Turbo-AI}. First of all, for a 2D massive array system, $\sum_{m=1}^M | \mathbf{W}_v^{(i,m)} |^2$ and $\sum_{n=1}^N | \mathbf{W}_h^{(j,n)} |^2 $ should be adjusted by selecting the number of antenna elements for both vertical and horizontal subspaces, or by configuring the antenna spacing to realize spatial correlation. Secondly, notice that the red shadowing part in Fig.~13 will be reduced, if $d =  \mathbf{W}_v^{(i,i)}\mathbf{W}_h^{(j,j)}$ increases. Once $d = 1$ holds, the shadowing part will disappear completely. This indicates that the subspace combining will not be effective any more, if the diagonal elements of matrices $\mathbf{W}_v$ and $\mathbf{W}_h$ become dominant, due to loss of spatial correlation. 

% Figure 13
\begin{figure}[!htp]
\vspace{-17mm}
\centering
\includegraphics[width=3.6in]{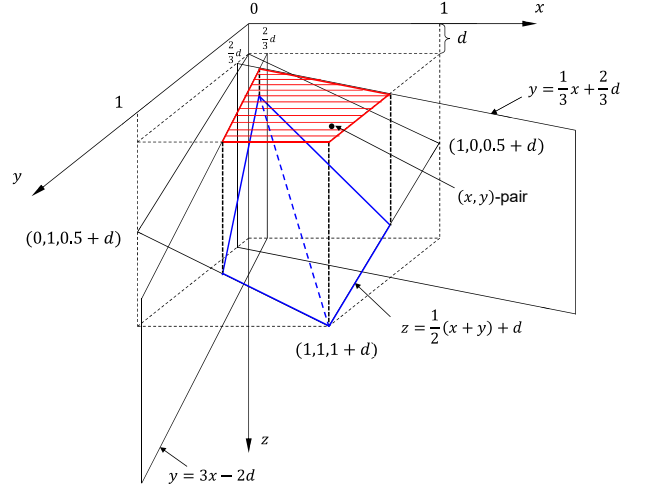}
\vspace{-41mm}
\caption{Effective combining region for $(x,y)$-pairs to boost the performance with \emph{Turbo-AI}}
\vspace{-2mm}
\end{figure}

\subsection{Proof of Turbo-AI}
\emph{Turbo-AI} is an iterative procedure. In training stage, \emph{Turbo-AI} exploits the new estimates to replace the observations and repeats the training. The applicability of \emph{Turbo-AI} comes from the fact that the additive noise variance after each training will be smaller than that of the previous observation, which can be represented by following equality,
% Equation 37
\vspace{1mm}
\begin{equation}
\label{eq37}
E\big[ \| \mathbf{W}_{I+1} \big(\prod_{i=0}^I \mathbf{W}_i \mathbf{Y} \big) - \mathbf{H} \|_F^2\big] \leq E\big[ \| \prod_{i=0}^I \mathbf{W}_i\mathbf{Y}-\mathbf{H} \|_F^2\big] \textrm{,}
\vspace{2mm}
\end{equation}where $\mathbf{W}_i$ denotes the weighting matrix of $i$-th iteration, $\mathbf{Y}$ and $\mathbf{H}$ denotes the observations and labels, respectively. Furthermore, the noise distribution at the input and output of the NN remains Gaussian due to the fact that our NN in essence is a linear operator on the input data $\mathbf{Y}$.

\begin{proof}
In \emph{Appendix A}, the convergence of a neural network is defined. Thus, after initializing \emph{Turbo-AI }at $0$-th iteration, if the convergence is reached after the learning stage, it holds,
% Equation 38
\vspace{1mm}
\begin{equation}
\label{eq38}
E\big[ \| \mathbf{W}_0 \mathbf{Y}  - \mathbf{H} \|_F^2\big] \leq E\big[ \| \underbrace{\mathbf{Y}-\mathbf{H}}_\mathbf{Z} \|_F^2\big] \textrm{.}
\vspace{2mm}
\end{equation}And at first iteration, as long as the convergence is reached with the updated observations $\mathbf{W}_0\mathbf{Y}$, a weighting matrix $\mathbf{W}_1$ exists, which satisfies the following inequality,
 % Equation 39
\vspace{1mm}
\begin{equation}
\label{eq39}
E\big[ \| \mathbf{W}_1 \mathbf{W}_0 \mathbf{Y}  - \mathbf{H} \|_F^2\big] \leq E\big[ \| \underbrace{\mathbf{W}_0\mathbf{Y}-\mathbf{H}}_{\mathbf{Z}_0} \|_F^2\big] \textrm{.}
\vspace{2mm}
\end{equation}
In order to prove (37) in sense of mathematical induction, let us assume that the following inequality holds at $I$-th iteration,
% Equation 40
\vspace{1mm}
\begin{equation}
\label{eq40}
E\big[ \| \mathbf{W}_{I} \big(\prod_{i=0}^{I-1} \mathbf{W}_i \mathbf{Y} \big) - \mathbf{H} \|_F^2\big] \leq E\big[ \| \underbrace{\prod_{i=0}^{I-1} \mathbf{W}_i\mathbf{Y}-\mathbf{H}}_{\mathbf{Z}_{I-1}} \|_F^2\big] \textrm{.}
\vspace{2mm}
\end{equation}Then, at $(I+1)$-th iteration, with the updated observations $\prod_{i=0}^{I}\mathbf{W}_i\mathbf{Y}$, the training can be carried out in the same neural network again. The weighting matrix $\mathbf{W}_{I+1}$ will be obtained, as the convergence is reached. It holds,
% Equation 41
\vspace{1mm}
\begin{equation}
\label{eq41}
E\big[ \| \mathbf{W}_{I+1} \big(\prod_{i=0}^I \mathbf{W}_i \mathbf{Y} \big) - \mathbf{H} \|_F^2\big] \leq E\big[ \| \underbrace{\prod_{i=0}^I \mathbf{W}_i\mathbf{Y}-\mathbf{H}}_{\mathbf{Z}_I} \|_F^2\big] \textrm{.}
\vspace{2mm}
\end{equation}
 (q.e.d.)   
\end{proof}

\vspace{1mm}
%\newpage 
%\section*{Acknowledgment}
%This work has been performed in the framework of the H2020 project 5GCAR co-funded by the EU. The views expressed are those of the authors and do not necessarily represent the project. The consortium is not liable for any use that may be made of any of the information contained therein.

%\vspace{1mm}

%\pagebreak
%\newpage

\end{document}